\documentclass[a4paper,11pt]{article}
\pdfoutput=1

\usepackage[utf8]{inputenc}
\usepackage{a4wide}
\usepackage{graphicx}
\usepackage{subcaption}
\graphicspath{{Figures/}} 
\usepackage{amssymb}
\usepackage{amsmath}
\usepackage{slashed}
\usepackage{mathtools}
\usepackage{cite} 
\usepackage{hyperref}
\usepackage{xcolor}
\usepackage{cancel}

\DeclareMathAlphabet{\mathpzc}{OT1}{pzc}{m}{it}

\newcommand{\be}{\begin{equation}}
\newcommand{\ee}{\end{equation}}
\newcommand{\ben}{\begin{eqnarray}}
\newcommand{\een}{\end{eqnarray}}
\newcommand{\bi}{\begin{itemize}}
\newcommand{\ei}{\end{itemize}}

\newcommand*\diff{\mathop{}\!\mathrm{d}}

\numberwithin{equation}{section}
\makeatletter
\g@addto@macro\bfseries{\boldmath}
\makeatother

\begin{document}

\begin{titlepage}
\begin{flushright}
IFT-UAM/CSIC-21-64
\end{flushright}
\vspace*{0.8cm}

\begin{center}
{\Large\bf Inverse Seesaw, dark matter and the Hubble tension}
\\[0.8cm]
E.~Fernandez-Martinez,$^a$
M.~Pierre,$^{a}$
E.~Pinsard,$^b$
and S.~Rosauro-Alcaraz$^a$\\[0.4cm]
$^{a}$\,{\it Departamento de F\'isica Te\'orica and Instituto de F\'{\i}sica Te\'orica, IFT-UAM/CSIC,\\
Universidad Aut\'onoma de Madrid, Cantoblanco, 28049, Madrid, Spain} \\
$^{b}$\,{\it Laboratoire de Physique de Clermont (UMR 6533), CNRS/IN2P3,
Univ. Clermont Auvergne, 4 Av. Blaise Pascal, F-63178 Aubière Cedex, France}\\
\end{center}
\vspace{0.8cm}

\begin{abstract}
We consider the inverse Seesaw scenario for neutrino masses with the approximate Lepton number symmetry broken dynamically by a scalar with Lepton number two. We show that the Majoron associated to the spontaneous symmetry breaking can alleviate the Hubble tension through its contribution to $\Delta N_\text{eff}$ and late decays to neutrinos. Among the additional fermionic states required for realizing the inverse Seesaw mechanism, sterile neutrinos at the keV-MeV scale can account for all the dark matter component of the Universe if produced via freeze-in from the decays of heavier degrees of freedom.
\end{abstract}
\end{titlepage}
\section{Introduction}
A significant and intriguing tension between local, late-time determinations of the Hubble rate and its preferred value when measured from early Universe probes persists. Analyses from type-Ia supernovae and strong lensing consistently favor values of $H_0$ significantly larger than those determined from the cosmic microwave background (CMB) and baryon acoustic oscillations data. A tension at the level of $4-6~\sigma$~\cite{Verde:2019ivm,Wong:2019kwg}, depending on the specific assumptions, exists between the {\tt Planck} value from the CMB spectrum~\cite{Aghanim:2018eyx} and the one obtained by the S$H_0$ES collaboration~\cite{Riess:2019cxk} from supernovae measurements. Among the many different solutions proposed~\cite{DiValentino:2021izs}, those that also address another open problem are particularly appealing. For instance, the authors of Refs.~\cite{Escudero:2019gvw,Escudero:2021rfi} have proposed that a light Majoron contributing to $\Delta N_\text{eff}$ and decaying to neutrinos after Big Bang Nucleosynthesis (BBN) may alleviate the discrepant determinations of $H_0$. This scenario would thus link the solution to the Hubble tension to the origin of neutrino masses and mixings.

Indeed, neutrino masses and mixings as required by the observation of the neutrino oscillation phenomenon, can be naturally accounted for through the Weinberg operator~\cite{Weinberg:1979sa}. This operator violates Lepton number ($L$) by two units. Thus, if this breaking is dynamical, the Majoron, the Goldstone boson associated to the spontaneous symmetry breaking of Lepton number, would be intimately linked to the origin of neutrino masses and mixings. 

Among the many possible realizations of the Weinberg operator, a particularly interesting one is the inverse Seesaw scenario~\cite{Mohapatra:1986aw,Mohapatra:1986bd,Bernabeu:1987gr}, which assumes an approximate Lepton number symmetry. In this way, the smallness of neutrino masses is naturally explained by this symmetry argument and a very high scale for the new physics is not required, leading to potentially more interesting phenomenological consequences.

In this work we investigate a dynamical origin for the small Lepton number breaking of the inverse Seesaw scenario.  Several constructions based on dynamical Lepton number breaking have been explored in the past 
\cite{Bazzocchi:2010dt,Dias:2011sq,DeRomeri:2017oxa,Bertuzzo:2018ftf,Ballett:2019cqp,Ballett:2019pyw,Gehrlein:2019iwl,Mandal:2020lhl,Mandal:2021acg}. In this work, we consider a Seesaw-like mechanism in the scalar sector so that a small vacuum expectation value is naturally induced for the scalar with Lepton number two~\cite{DeRomeri:2017oxa}. Thus, neutrino masses will be proportional to this small parameter. The dynamical symmetry breaking will also imply the presence of a Majoron with the potential of alleviating the Hubble tension. 

Furthermore, inverse Seesaw realizations may also lead to sterile neutrinos at the few keV scale which are good dark matter (DM) candidates~\cite{Abada:2014zra,Abada:2014vea,Boulebnane:2017fxw}. While production via mixing~\cite{Dodelson:1993je} is ruled out\footnote{Except in the presence of a sizable Lepton number asymmetry~\cite{Shi:1998km,Ghiglieri:2019kbw,Ghiglieri:2020ulj}.} by the stringent constraints from searches of X-ray lines~\cite{Roach:2019ctw}, the correct DM relic abundance could be achieved from the decay of heavier states instead~\cite{Petraki:2007gq,Boulebnane:2017fxw,Merle:2013wta, Abada:2014vea, Abada:2014zra, Drewes:2015eoa, Adhikari:2016bei, DeRomeri:2020wng}. After investigating this possibility, we find that the couplings of the scalars to neutrinos can lead to both the correct DM relic abundance through freeze-in via decays and the necessary Majoron population so as to alleviate the Hubble tension. 

This paper is organized as follows, in Section~\ref{sec:model} we introduce the particle content and Lagrangian of the inverse Seesaw model with dynamical breaking of $L$ considered. In Section~\ref{sec:DM} we discuss the dark matter production mechanism as well as its phenomenology and constraints in the parameter space. In Section~\ref{sec:H0} we analyze the conditions under which also the Hubble tension can be alleviated as proposed in Refs.~\cite{Escudero:2019gvw,Escudero:2021rfi}. Finally, in Section~\ref{sec:summary} we discuss and summarize the allowed regions of the parameter space while in Section~\ref{sec:conclusions} we conclude.

\section{The model}
\label{sec:model}

The simplest extension of the Standard Model (SM) particle content to account for the neutrino masses and mixing evidence is the addition of fermion singlets, \textit{i.e.} right-handed neutrinos. Furthermore, if a large Majorana mass is also included for the right-handed neutrinos, the smallness of neutrino masses is naturally explained through the hierarchy between this mass and the electroweak scale via the celebrated canonical type-I Seesaw~\cite{Minkowski:1977sc,Mohapatra:1979ia,Yanagida:1979as,GellMann:1980vs}. Conversely, a Majorana mass significantly above the electroweak scale destabilizes the Higgs mass, worsening the electroweak hierarchy problem~\cite{Vissani:1997ys,Casas:2004gh} and greatly hinders the testability of the mechanism. 

It is therefore appealing to consider low-scale alternatives to the canonical Seesaw mechanism. This option was investigated in Refs.~\cite{Branco:1988ex,Kersten:2007vk, Abada:2007ux,Moffat:2017feq} showing that the smallness of neutrino masses can also be naturally explained by an approximate Lepton number symmetry. Two types of fermion singlets may be included according to their $L$ assignment. The first option is Dirac pairs with $L=1$ for which, in the limit of exact $L$, only the right-handed component may have a Yukawa coupling to the active SM neutrinos. The second option is Majorana sterile neutrinos with $L=0$ which, for exact $L$ symmetry, do not couple to any other fermion. At this level, three neutrinos remain massless. When the $L$ symmetry is slightly broken, small neutrino masses can be induced, the Dirac neutrinos may split into pseudo-Dirac pairs, and additional suppressed couplings are allowed.

Following this principle, we extend the SM particle content with Dirac pairs with their corresponding right and left-handed components $N_R$ and $N_L$. At least two of these pairs are required to reproduce the correct neutrino masses and mixings (in this case with the lightest neutrino massless). We also consider one Majorana sterile neutrino $n_L$. The $L$ violation that will induce the standard neutrino masses will be dynamical and originated through two scalars $\phi_1$ and $\phi_2$. All the new states are singlets of the SM gauge group and their Lepton number charge  assignment $L$ is given in Table~\ref{Tab:Charge_assignment1}. 

\begin{table}[h]
\centering
 \begin{tabular}{|c|c|c|c|c|c|} 
 \hline
 State & $N_R$ & $N_L$ & $n_L$ & $\phi_1$ & $\phi_2$\\
 \hline
 $L$ & 1 & 1 & 0 & 1 & 2 \\
 \hline
 \end{tabular}
 \caption{New fermions and scalars with their charge under Lepton number. $\phi_{1,2}$ are SM singlet scalars while $N_R$ is right-handed and $N_L$ and $n_L$ are left-handed SM singlet fermions.}
 \label{Tab:Charge_assignment1}
\end{table}

According to this assignment the Lagrangian of the model can be parametrized as:
\begin{equation}
\begin{split}
    \mathcal{L}\supset& -\overline{L}_L\tilde{H}Y_{\nu}N_R-\overline{N}_LMN_R-\frac{1}{2}\overline{N}_L\phi_2 Y_{LL} N_L^c-\frac{1}{2}\overline{N_R^c}\phi_2Y_{RR}N_R\\
    &-\frac{1}{2}\overline{n}_L\mu_{ss}n_L^c-\overline{N}_L \phi_1 Y_{Ls} n_L^c-\overline{N^c_R}\phi_1^\dagger Y_{Rs} n_L^c+\text{h.c.}+V(\phi_1,\phi_2,H),
    \label{Eq:Lag}
\end{split}
\end{equation}
where $L_L$ are the SM lepton doublets and $H$ the Higgs doublet. 

\subsection{Scalar potential}

In the inverse Seesaw mechanism, neutrino masses are protected by an approximate $L$ symmetry. The smallness of the $L$-violating terms thus naturally explains the lightness of neutrino masses. Since we want to explore a dynamical breaking of this symmetry, we also consider a Seesaw-like mechanism in the scalar sector to avoid hierarchy problems and account for the smallness of the $L$-violating vacuum expectation value (vev) in a technically natural way. To this end, we mimic the type-II Seesaw~\cite{Magg:1980ut,Schechter:1980gr,Lazarides:1980nt,Mohapatra:1980yp} and assume that the vev of $\phi_2$ will be induced by that of $\phi_1$ as in Ref.~\cite{DeRomeri:2017oxa}. In particular, the scalar potential is given by
\begin{equation}
\begin{split}
    V\,=\,&\frac{m_H^2}{2}H^{\dagger}H+\frac{\lambda_H}{2}(H^{\dagger}H)^2+\frac{m_1^2}{2}\phi_1^*\phi_1+\frac{m_2^2}{2}\phi_2^*\phi_2+\frac{\lambda_1}{2}(\phi_1^*\phi_1)^2+\frac{\lambda_2}{2}(\phi_2^*\phi_2)^2\\
    +&\frac{\lambda_{1H}}{2}(\phi_1^*\phi_1)(H^{\dagger}H)+\frac{\lambda_{2H}}{2}(\phi_2^*\phi_2)(H^{\dagger}H)+\frac{\lambda_{12}}{2}(\phi_1^*\phi_1)(\phi_2^*\phi_2)-\eta(\phi_1^2\phi_2^*+\phi_1^{*2}\phi_2).
    \label{Eq:Scalar_potential}
\end{split}
\end{equation}
If both $m_H^2$ and $m_1^2$ are negative but $m_2^2$ is positive and large, then the vev of $\phi_2$, $v_2$, is only induced by the vev of $\phi_1$, $v_1$, through $\eta$ and can be made naturally small. Indeed, notice that in the limit $\eta \rightarrow 0$ together with $Y_{Ls} \rightarrow 0$ and $Y_{Rs} \rightarrow 0$, the Lagrangian would be invariant under a separate $U(1)$ transformation of $\phi_1$, different from $L$. Thus, these three parameters are protected by an additional symmetry and very small values for them are natural in the 't~Hooft sense. Parametrising the scalars as $\phi_i=(v_i+\varphi_i)e^{i a_i/v_i}/\sqrt{2}$ and $H=(v_H+h)/\sqrt{2}$ in the unitary gauge, the minimisation conditions read as
\begin{equation}
    \begin{gathered}
        m_H^2\,=\,-\frac{1}{2}\left(2\lambda_H v_H^2+\lambda_{1H}v_1^2+\lambda_{2H}v_2^2\right)\simeq-\frac{1}{2}\left(2\lambda_H v_H^2+\lambda_{1H}v_1^2\right)\,,\\
        m_1^2\,=\,-\dfrac{1}{2} \left(2 \lambda_1 v_1^2+\lambda_{1H} v_h^2 +  \lambda_{12} v_2^2 -4 \sqrt{2} \eta  v_2  \right)\simeq-\frac{1}{2}\left(2\lambda_1v_1^2+\lambda_{1H}v_H^2\right)\,,\\       
        m_2^2\,=\,-\frac{1}{2}\left(2\lambda_2 v_2^2+\lambda_{12}v_1^2+\lambda_{2H}v_H^2-\frac{2 \sqrt{2} \eta v_1^2}{v_2}\right)\simeq \frac{\sqrt{2} \eta v_1^2}{v_2},
    \end{gathered}
    \label{Eq:Minimisation}
\end{equation}
and thus
\begin{equation}
    v_2 \simeq \frac{\sqrt{2} \eta v^2_1}{m^2_2},
    \label{eq:v2}
\end{equation}
where we have assumed $v_2\ll v_1,v_H$. From Eq.~(\ref{eq:v2}) we can see that indeed $v_2$ is induced from the vev $v_1$ and suppressed by $\eta$ so that $v_2\rightarrow 0$ if $\eta \rightarrow 0$ or $m_2\rightarrow \infty$. The scalar mass matrix in the basis $\begin{pmatrix} h & \varphi_1 & \varphi_2 \end{pmatrix}$, in the $v_2\rightarrow 0$ limit reads
\begin{equation}
    M^2\simeq\begin{pmatrix}
    \lambda_H v_H^2 & \frac{1}{2}\lambda_{1H}v_1 v_H & 0\\
    \frac{1}{2}\lambda_{1H}v_1 v_H & \lambda_1 v_1^2 & -\sqrt{2}\eta v_1\\
    0 & -\sqrt{2}\eta v_1 & \frac{\eta v_1^2}{\sqrt{2}v_2}
    \end{pmatrix},
\end{equation}
so that the masses of the physical scalars $h$, $\varphi_1$ and $\varphi_2$ are approximately\footnote{We use the same notation for the mass and flavour CP-even scalar eigenstates for brevity as mixing angles are typically small.}
\begin{equation}
    m_h^2\simeq \lambda_H v_H^2,\quad m_{\varphi_1}^2\simeq \lambda_1 v_1^2,\quad m_{\varphi_2}^2\simeq m_2^2/2\,,
\end{equation}
for small mixed quartic couplings. The mixing angles $\alpha_{1H}$ and $\alpha_{12}$ between $h-\varphi_1$ and $\varphi_1-\varphi_2$ are, respectively,
\begin{equation}
    \tan{\big(2\alpha_{1H}\big)}\simeq - \frac{\lambda_{1H}v_1 v_H}{\lambda_1 v_1^2-\lambda_H v_H^2}\,,\quad \text{and}\quad \tan{\big(2\alpha_{12}\big)}\simeq 4\frac{v_2}{v_1}\,.
\end{equation}
The physical pseudoscalars are given by
\begin{equation}
\begin{gathered}
    J=\frac{1}{\sqrt{v_1^2+4v_2^2}}\left(v_1 a_1+2 v_2 a_2\right),\quad m_J^2 = 0\,,\\
    A=\frac{1}{\sqrt{v_1^2+4v_2^2}}\left(-2v_2 a_1+ v_1 a_2\right),\quad m_A^2 \simeq m_{\varphi_2}^2\,,
\end{gathered}
\end{equation}
where $J$ is the Goldstone boson associated to the breaking of $L$, that is, the Majoron, and therefore massless from the scalar potential. Since $L$ is expected to be broken from gravity effects~\cite{Akhmedov:1992hi}, we will assume that a Majoron mass of the order of the eV scale is induced by them. The mixing angle $\beta$ between $a_1-a_2$ is:
\begin{equation}
    \tan{\big(2\beta\big)}\simeq 4\frac{v_2}{v_1}.
\end{equation}

\subsection{Neutrino masses}

When all the scalars develop their vevs, $v_H$, $v_1$ and $v_2$ respectively, the neutrino mass matrix takes the inverse Seesaw form:

\begin{equation}
    \mathcal{M}=\begin{pmatrix}
    0 & 0 & 0 & m_D\\
    0 & \mu_{ss} & \mu_{Ls}^T & \mu_{Rs}^T\\
    0 & \mu_{Ls} & \mu_{LL} & M^T\\
    m_D^T & \mu_{Rs} & M & \mu_{RR}
    \end{pmatrix},
\end{equation}
where we have defined $m_D\equiv v_HY_{\nu}/\sqrt{2} $, $\mu_{LL}\equiv v_2Y_{LL}/\sqrt{2}$,  $\mu_{RR}\equiv v_2Y_{RR}/\sqrt{2}$, $\mu_{Ls}\equiv v_1Y_{Ls}/\sqrt{2}$ and $\mu_{Rs}\equiv v_1Y_{Rs}/\sqrt{2}$, arranging the states as $\begin{pmatrix} \nu_L & n_L & N_L & N_R^c\end{pmatrix}$. From now on we will work in the basis where $M$ is diagonal. The approximate expressions for the flavour states in terms of the mass eigenstates are:
\begin{equation}
        \begin{gathered}
        \nu_L\simeq U\nu_i-\theta^*\left(\mu_{Ls}^*\mu_{ss}^{-1}+M^{-1}\mu_{Rs}\right)\nu_4+\frac{1}{\sqrt{2}}\theta^*\left(N_+-iN_-\right),\\
        n_L\simeq \mu_{ss}^{-1}\mu_{Ls}^T\theta^TU\nu_i+\nu_4+\frac{1}{\sqrt{2}}\mu_{Ls}^\dagger M^{-1}\left(N_++iN_-\right)+\frac{1}{\sqrt{2}}\mu_{Rs}^\dagger M^{-1}\left(N_+-iN_-\right),\\
        N_L\simeq -\theta^TU \nu_i+\left(\theta^T\theta^*\mu_{Ls}^*\mu_{ss}^{-1}-M^{-1}\mu_{Rs}\right)\nu_4+\frac{1}{\sqrt{2}}\left(N_+-iN_-\right),\\
        N_R^c\simeq -M^{-1}\mu_{Ls}\nu_4+\frac{1}{\sqrt{2}}\left(N_++iN_-\right),
    \end{gathered}
    \label{Eq:Diagonalization_Final1}
\end{equation}
where $\nu_i$, $\nu_4$, $N_+$ and $N_-$ are the mass eigenstates with masses
\begin{equation}
    \begin{gathered}
    m_{\nu_i}\simeq U^T \theta\left(\mu_{LL}-\mu_{Ls}\mu_{ss}^{-1}\mu_{Ls}^T\right)\theta^T U,\quad m_{\nu_4}\simeq \mu_{ss}, \\
    m_{N_\pm}\simeq \sqrt{M^2+m_D^{\dagger}m_D}\pm\frac{1}{2}\left(\mu_{LL}+\mu_{RR}\right),
    \end{gathered}
    \label{eq:nu_masses}
\end{equation}
with $i=1,2,3$, $\theta\equiv m_DM^{-1}$ characterizing the mixing between the active flavours $\nu_L$ and the heavy states $N_\pm$, and $U$ the unitary matrix diagonalising the light neutrino mass matrix after the block diagonalisation. We have assumed that $M \gg m_D \gg \mu$. In particular, we will assume that $M$ is somewhat above the electroweak scale and that it controls the scale of the pseudo-Dirac pairs $N_\pm$. The splitting of the pseudo-Dirac pairs is only through the Majorana masses $\mu_{LL}$ and $\mu_{RR}$. We will also assume that $\mu_{ss}$ is at the keV scale and is the main contribution to $m_{\nu_4}$, the dark matter candidate mass. For a summary of the approximate ranges of all the model parameters to correctly reproduce neutrino masses and mixings, the DM relic abundace and to improve on the Hubble tension see Table~\ref{tab:parameters} where we sumarize our findings of the following sections. According to these values, all the $\mu$ parameters have been treated as a perturbation in the expressions above and the results are to leading order in perturbation theory. Furthermore, we have also approximated the results to leading order in $\theta$ to simplify the expressions. Notice that $U$, the rotation diagonalising the light neutrino mass matrix, $m_{\nu_i}$, corresponds to the PMNS mixing matrix at leading order. 

\section{Dark matter}
\label{sec:DM}

The Majorana fermion singlet $n_L$, with its $L=0$ charge assignment, can only mix with the other neutrinos via $L$-violating, and therefore suppressed, parameters. Furthermore, its allowed interactions with $N_L$ and $N_R$ are via $\phi_1$ through the $Y_{Ls}$ and $Y_{Rs}$ parameters respectively. These two parameters, together with $\eta$, are all protected by an additional symmetry. Indeed, setting the three of them to zero a new $U(1)$ transformation for $\phi_1$, independent from $L$, becomes a symmetry of the Lagrangian. We will therefore consistently assume small values for these three parameters. As previously discussed, a small value of $\eta$ guarantees that the induced vev $v_2$ will be suppressed and thus naturally explain the smallness of neutrino masses. Small values for $Y_{Ls}$ and $Y_{Rs}$ in turn imply that interactions and decays of $n_L$ are very suppressed, making it an ideal DM candidate via freeze-in production. In this way, the same symmetry behind the smallness of neutrino masses also guarantees DM stability in a natural way.  In the following we will discuss the production mechanism as well as the main constraints from it and other observations on the parameter space of the model.  

\subsection{Dark matter production}

The DM candidate in our model is the mass eigenstate $\nu_4$ which is approximately aligned with the fermion singlet $n_L$ with only suppressed mixings with the rest of the interaction eigenstates given the approximate $L$ symmetry, as shown in Eq.~(\ref{Eq:Diagonalization_Final1}). While the active flavour eigenstates $\nu_L$ do contain an admixture of the DM candidate $\nu_4$ as given by Eq.~(\ref{Eq:Diagonalization_Final1}), processes that produce $\nu_L$ such as decays of the $Z$ and $W$ or of the heavy neutrinos $N_\pm$ via their Yukawa interactions with the Higgs, will not contribute to the production of $\nu_4$ beyond the standard Dodelson-Widrow mechanism~\cite{Dodelson:1993je}. Indeed, the active flavour eigenstates $\nu_L$ are already in thermal equilibrium in the early Universe and additional contributions such as these will not modify their abundance. In other words, the thermal masses of the active neutrinos $\nu_L$ are very relevant in the early Universe and dominate over the keV-scale mass of $\nu_4$, suppressing the mixing~\cite{Lello:2016rvl}. That is, the interaction eigenstates are approximately the effective mass eigenstates~\cite{Abada:2018oly}. Therefore, in this regime, it is more convenient to work in a mixed basis with $N_\pm$ together with $\hat{\nu}_L$ and $\hat{n}_L$: the ``incomplete'' flavour states $\nu_L$ and $n_L$ at energies below $m_{N_\pm}$. In this intermediate basis the original interaction eigenstates read: 

\begin{equation}
\begin{gathered}
    \nu_L\simeq \hat{\nu}_L+\frac{\theta^*}{\sqrt{2}}(N_+-iN_-),\\
    n_L\simeq\hat{n}_L+\frac{1}{\sqrt{2}}\mu_{Ls}^\dagger M^{-1}\left(N_++iN_-\right)+\frac{1}{\sqrt{2}}\mu_{Rs}^\dagger M^{-1}\left(N_+-iN_-\right),\\
    N_L\simeq-\theta^T \hat{\nu}_L-M^{-1}\mu_{Rs}\hat{n}_L+\frac{1}{\sqrt{2}}(N_+-iN_-),\\
    N_R^c\simeq-M^{-1}\mu_{Ls}\hat{n}_L+\frac{1}{\sqrt{2}}(N_++iN_-).
\end{gathered}
\end{equation}
Since $\hat{n}_L$ does not share the relevant contributions to the thermal masses with $\hat{\nu}_L$, it is through processes in which $\hat{n}_L$ is produced where contributions to the final DM abundance of $\nu_4$ beyond the Dodelson-Widrow mechanism can be achieved. The main interactions of $\hat{n}_L$ are with the heavy pseudo-Dirac pairs $N_\pm$ and the new scalar particles via the couplings $Y_{Ls}$ and $Y_{Rs}$. Thus, given the smallness of $Y_{Ls}$ and $Y_{Rs}$, the main production channel for DM is through freeze-in~\cite{Hall:2009bx} decays of the heavy neutrinos to a DM state and $S$, with $S=\varphi_{1(2)},A,J$ any of the physical scalar degrees of freedom. We find that the total dark matter production rate is
\begin{align}
    \Gamma_{\hat{n}_L}=\,&\sum_{i=\pm}\Gamma\left(N_i\rightarrow \varphi_2+\hat{n}_L\right)+ \Gamma\left(N_i\rightarrow A+\hat{n}_L\right)+\Gamma\left(N_i\rightarrow \varphi_1+\hat{n}_L\right)+ \Gamma\left(N_i\rightarrow J+\hat{n}_L\right) \,,\nonumber \\
    =\,&\frac{m_N}{16\pi}\Bigg\lbrace \left(\frac{\mu_{Ls}}{v_1}\right)^2\left(1+\frac{\mu_{Rs}^2}{\mu_{Ls}^2}\right)\Bigg[c_{\beta}^2+c_{12}^2\left(1-\frac{m_{\varphi_1}^2}{m_N^2}\right)^2\Theta(m_N-m_{\varphi_1})\Bigg] +\left(\frac{\mu_{LL}\mu_{Rs}}{v_2M}\right)^2 \,\nonumber \\
    & \times \left(1+\frac{\mu_{RR}^2\mu_{Ls}^2}{\mu_{LL}^2\mu_{Rs}^2}\right) \Bigg[c_{12}^2\left(1-\frac{m_{\varphi_2}^2}{m_N^2}\right)^2\Theta(m_N-m_{\varphi_2})
    +c_{\beta}^2\left(1-\frac{m_{A}^2}{m_N^2}\right)^2\Theta(m_N-m_{A})\Bigg]\Bigg\rbrace,
\end{align}
 where $\Theta$ is the Heaviside step function and $c_{12}\equiv \cos{\alpha_{12}}$ and $c_{\beta}\equiv \cos{\beta}$. We have made the approximation $m_{N_+}\sim m_{N_-}\equiv m_N$. If the heavy neutrinos thermalize with the SM plasma, which happens for all the values of $\theta$ we will consider, the relic density is given by
\begin{align}
        \Omega_{\nu_4} h^2 \, \simeq  \, m_{\nu_4} M_P  \sqrt{\frac{5}{\pi}} \frac{405  }{8 \pi^4 g_{\star}^{3/2}(m_N) }  \dfrac{\Gamma_{\hat{n}_L}}{m_N^2} \dfrac{s^0}{\rho_{c}^0} h^2 \,,
\end{align}
where $M_P=1.22 \cdot 10^{19}$~GeV is the Planck mass, $s^0$ and $\rho^0_c$ are the present entropy density and critical energy density respectively, $h$ is the present Hubble constant expressed in units of $100\, \text{km}\,\text{s}^{-1}\text{Mpc}^{-1}$ and $g_{\star}(m_N)$ is the number of radiation degrees of freedom during the $N_i$ decays, which we approximate as $g_{\star}(m_N) = 106.75$ for $m_N \gtrsim 100\,\text{GeV}$.
This expression can be simplified to study the analytical scaling of the relic density in two opposing limits of the following ratio
\begin{equation}
    r\,\equiv \, \dfrac{\mu_{Rs}}{\mu_{Ls}}\,.
\end{equation}
Nevertheless the full expression of the relic density has been taken into account in the numerical results. Assuming small $r\ll1$ and that the decay width is dominated by the $J$ and  $\varphi_1$ final states, the relic abundance scales as
\begin{equation}
    \Omega_{\nu_4} h^2 (r\ll 1) \, \simeq \, 0.13 \,\left( \dfrac{m_{\nu_4}}{10\,\text{keV}} \right)^3 \left( \dfrac{U_{\alpha4}}{ 10^{-6}} \right)^2  \left( \dfrac{10^{-5}}{\theta} \right)^2 \left( \dfrac{150\,\text{GeV}}{m_{N}} \right) \left( \dfrac{200\,\text{GeV}}{v_1} \right)^2\,,
\end{equation}
where we have neglected the scalar masses with respect to $m_N$  and approximated $c_\beta \sim c_{12} \sim 1$ as well as written $\mu_{Ls}$ in terms of $U_{\alpha4} \sim \theta \mu_{Ls}/m_{\nu_4}$,
the mixing between the DM candidate $\nu_4$ and the active neutrinos $\nu_{L_\alpha}$ with $\alpha= e, \mu, \tau$ (see Eq.~(\ref{Eq:Diagonalization_Final1})) for which strong constraints exist from X-ray searches. 

In the opposite limit $r\gg 1$ and assuming dominant decays to $\varphi_2$ and $A$, again neglecting their masses with respect to $m_N$, the relic density scales approximately as
\begin{equation}
    \Omega_{\nu_4} h^2 (r\gg 1)  \simeq  0.11 \left( \dfrac{r}{10^{3}} \right)^2\left( \dfrac{m_{\nu_4}}{10\,\text{keV}} \right)^3 \left( \dfrac{m_{\nu_i}}{0.05\,\text{eV}} \right)^2 \left( \dfrac{U_{\alpha4}}{ 10^{-6}} \right)^2  \left( \dfrac{10^{-3}}{\theta} \right)^6 \left( \dfrac{120\,\text{GeV}}{m_{N}} \right)^3 \left( \dfrac{\text{MeV}}{v_2} \right)^2
\end{equation}
\begin{figure*}[t!]
\centering
\begin{subfigure}[t]{0.495\textwidth}
\centering
\includegraphics[width=\textwidth]{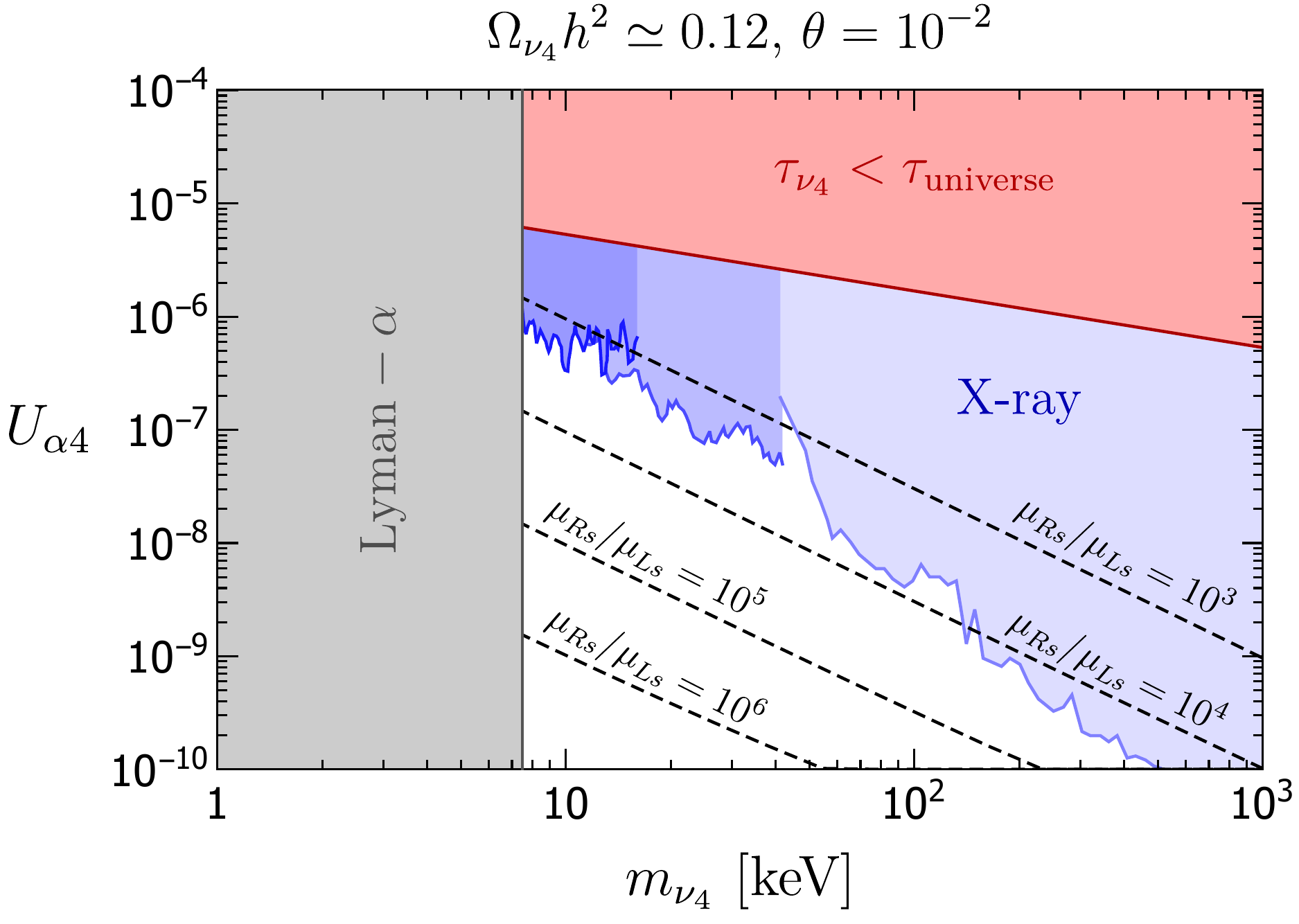}
\label{fig:U41_mnu4_theta-2}
\end{subfigure}
\hfill
\begin{subfigure}[t]{0.495\textwidth}  
\centering 
\includegraphics[width=\textwidth]{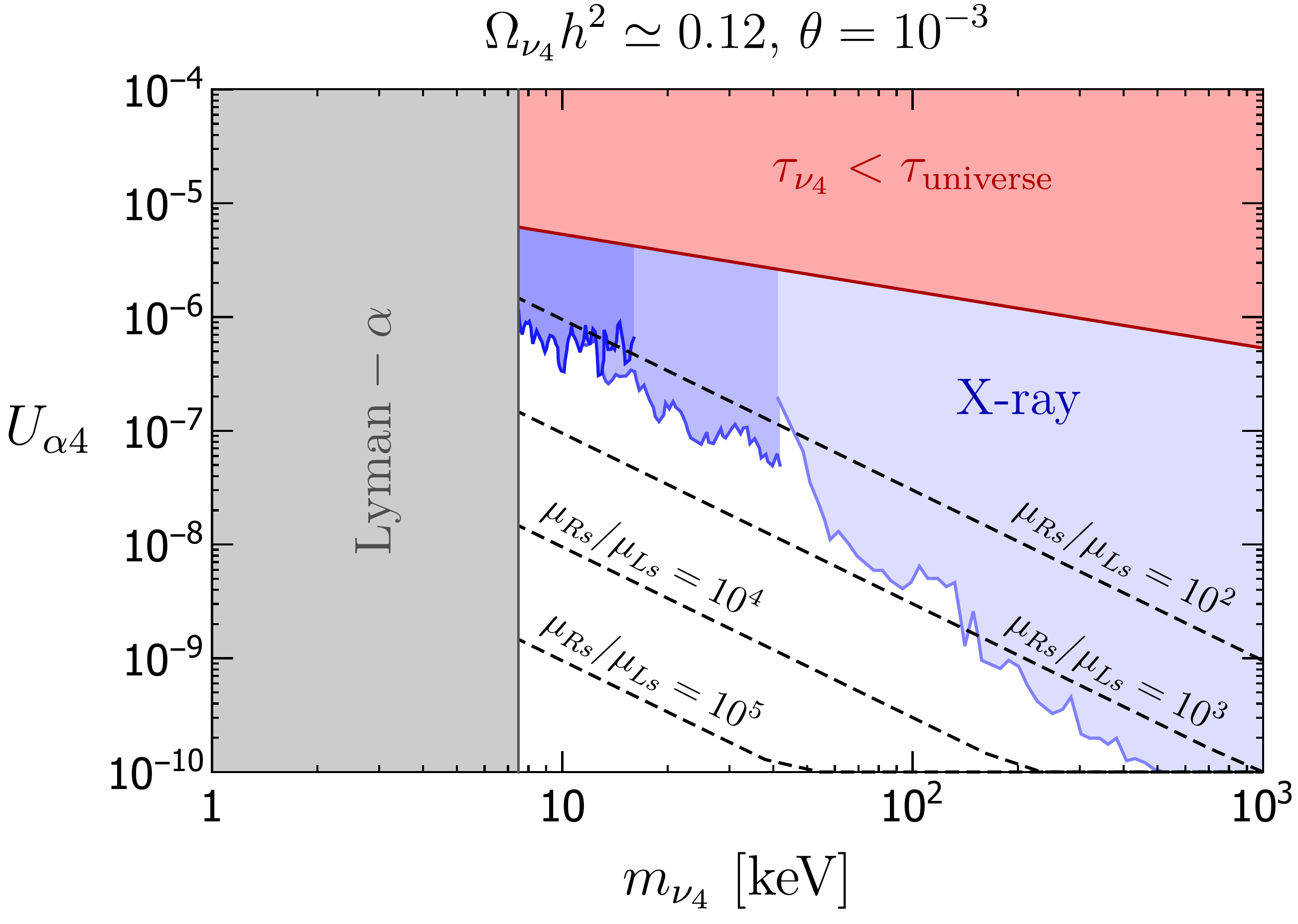}
\label{fig:U41_mnu4_theta-3}
\end{subfigure}
\caption[]
{\small  Available parameter space allowing to reproduce the correct DM relic abundance for $\theta=10^{-2},10^{-3}$. The y-axis is the mixing angle between the dark matter mass eigenstate and the SM neutrino flavour eigenstates and the x-axis the dark matter mass.
Black dashed lines represents the correct relic density for a fixed ratio $\mu_{Rs}/\mu_{Ls}$. The three blue shaded regions dubbed ``X-ray'' represent respectively, from left to right, constraints from  XMM-Newton \cite{Foster:2021ngm}, NuSTAR \cite{Roach:2019ctw} and INTEGRAL \cite{Boyarsky:2007ge} on spectral photon lines generated by decaying dark matter. The grey region corresponds to Lyman-$\alpha$ constraints from the matter power spectrum on light free-streaming dark matter particles, estimated from \cite{Ballesteros:2020adh}. The red region represents the parameter space for which the dark matter lifetime is shorter than the age of the Universe, estimated using Eq. (\ref{eq:DMlifetime}). In this figure we fixed $m_N=150$  GeV, $m_{\varphi_2}=m_A=50$ GeV, $v_1=200$ GeV, $m_{\varphi_1}\ll m_N$, $v_2=1$ MeV, $\mu_{RR}=10$ keV.}
\label{fig:U41_mnu4}
\end{figure*} 
In Fig.~\ref{fig:U41_mnu4} we show the allowed regions of the parameter space allowed by the X-ray searches and Lyman-$\alpha$ forest constraints on neutrino DM as well as the values of $\mu_{Rs}/\mu_{Ls}$ for which the correct relic abundance would be obtained for different values of $\theta$.
The parameter $\theta$ represents the mixing between the active neutrinos and the heaviest mass eigenstates and can be bounded to be $\theta \leq 10^{-2}$ from flavour and electroweak precision tests of the unitarity of the PMNS matrix~\cite{Fernandez-Martinez:2016lgt}. As can be seen, for values of $\theta$ close to the current upper bound in Fig.~\ref{fig:U41_mnu4}, a sizable hierarchy of about four orders of magnitude between $\mu_{Rs}$ and $\mu_{Ls}$ would be needed in order to obtain the correct relic abundance through $\mu_{Rs}$. Indeed, the stringent X-ray constraints require sufficiently suppressed active neutrino-DM mixing, which is mainly\footnote{Notice that from Eq.~(\ref{Eq:Diagonalization_Final1}) there is also a contribution to $U_{\alpha4}$ from $\mu_{Rs}$. This contribution is suppressed by the ratio $\sim m_{\nu_4}/m_N$ with respect to the one from $\mu_{Ls}$. Therefore, the dominant contribution is always $\mu_{Ls}$ for the parameter space shown in Fig.~\ref{fig:U41_mnu4} even for the largest values of $r$ depicted.} dominated by $\mu_{Ls}$. Conversely, this hierarchy is avoided for smaller values of $\theta$. This choice may be considered more natural, since there is no reason for a significant hierarchy among these parameters from the charge assignments of the fields. However, a lower bound on $\theta$ can be extrated from the requirement of perturbative unitarity for the Yukawa coupling $Y_{LL}$ by the relation
\begin{equation}
    \theta\,\simeq\,2.5\cdot 10^{-4}\, \left( \dfrac{1}{Y_{LL}}\right)^{1/2}    \left( \dfrac{m_{\nu_i}}{0.05 \, \text{eV}}\right)^{1/2}\left( \dfrac{\text{MeV}}{v_2}\right)^{1/2}\,,
\end{equation}
implying that for $v_2 \lesssim 1$ GeV, $\theta$ has to be larger than $\mathcal{O}(10^{-5})$ to ensure perturbativity for $Y_{LL}$. Moreover, small values of $\theta$ would reduce the testability of this region of the parameter space, at least through unitarity constraints of the PMNS matrix or direct searches of the heavy neutrinos at colliders. In this regime, the dominant phenomenology of the model would rather correspond to DM searches via X-rays as well as through cosmology from its impact on the $H_0$ tension and contributions to $\Delta N_\text{eff}$.

\subsection{Dark matter decay $\nu_4 \rightarrow \nu_i + \gamma$}
Sterile-neutrino like dark matter can decay into a neutrino and a photon producing a monochromatic spectral line. The dark matter mixing with active neutrinos, as given by Eq.~(\ref{Eq:Diagonalization_Final1}), is constrained by the International Gamma-Ray Astrophysics Laboratory (INTEGRAL) \cite{Boyarsky:2007ge} by looking for DM decaying in the Milky Way halo, as well as from NuSTAR \cite{Roach:2019ctw} and XMM-Newton \cite{Foster:2021ngm}. These constraints correspond to the blue regions in Fig.~\ref{fig:U41_mnu4}.

\subsection{Dark Matter lifetime}
Notice that, apart from the usual decay channels to three light neutrinos or a neutrino and an X-ray photon, DM may also decay to a Majoron and a light neutrino. Thus, the associated lifetime of the DM needs to be larger than the age of the Universe. The decay rate is given by
\begin{equation}
    \Gamma\left(\nu_4\rightarrow J+\nu_i\right)=\frac{m_{\nu_4}}{16\pi}s_{\beta}^2\left(\theta^3\frac{\mu_{Ls}}{\mu_{ss}}\frac{\mu_{LL}}{v_2}\right)^2,
\end{equation}
 which gives
\begin{equation}
  \dfrac{\Gamma\left(\nu_4\rightarrow J+\nu_i\right)^{-1}}{ \tau_\text{universe}} \,\simeq\,28
    \, \left( \dfrac{v_1}{200\;\text{GeV}} \right)^2 \left( \dfrac{10\;\text{keV}}{m_{\nu_4}} \right) \left( \dfrac{0.05\;\text{eV}}{m_{\nu_i}} \right)^2 \left( \dfrac{10^{-6}}{U_{\alpha4}} \right)^2\,.
    \label{eq:DMlifetime}
\end{equation}
The stability condition $\tau_{\nu_4}  \equiv \Gamma\left(\nu_4\rightarrow J+\nu_i\right)^{-1} > \tau_\text{universe} $ excludes the parameter space corresponding to the red region depicted in Fig.~\ref{fig:U41_mnu4}. Nonetheless, notice that the constraints on the mixing from X-ray searches are always stronger than those from the DM lifetime.

\subsection{Constraints from the power spectrum and  Lyman-$\alpha$}

Light dark matter candidates carrying a non-negligible amount of kinetic energy can alter $\Lambda$CDM predictions of the matter power spectrum which are probed on the smallest physical scales, i.e. largest Fourier wavenumbers $k\sim (0.1-10)\, h\,\text{Mpc}^{-1}$, by the so-called Lyman-$\alpha$ forest. Constraints from Lyman-$\alpha$ on such DM candidates are typically given in terms of a lower bound for the  Warm Dark Matter (WDM) mass~\cite{Narayanan:2000tp,Viel:2005qj,Viel:2013fqw,Baur:2015jsy,Irsic:2017ixq,Palanque-Delabrouille:2019iyz,Garzilli:2019qki}

\begin{equation}
m_{\rm WDM}\;\gtrsim\; m_{\rm WDM}^{\text{Ly}-\alpha}\; =\; (1.9-5.3)~\text{keV at 95\% C.L.} \,,
\label{eq:bound_MWDM_ly-alpha}
\end{equation}
In our scenario the DM density is generated from the decay of non-relativistic heavy neutrinos thermalized with the SM plasma. The effect of the resulting non-thermal phase space distribution on the matter power spectrum has been studied in various works \cite{Petraki:2007gq,Boulebnane:2017fxw,Bae:2017dpt,Kamada:2019kpe,Heeck:2017xbu,DEramo:2020gpr}. The Lyman-$\alpha$ constraints on our DM candidate can be expressed, following the procedure of~\cite{Ballesteros:2020adh}, as 
\begin{equation}
       m_{\nu_4} \, \gtrsim \, 7.5 ~\text{keV}\,  \left( \dfrac{m_{\rm WDM}^{\text{Ly}-\alpha}}{3~\text{keV}} \right)^{4/3} \left( \dfrac{106.75}{g_{\star s}(m_N)} \right)^{1/3}\,,
       \label{eq:bound_ly-alpha}
\end{equation}
where $g_{\star s}(T)$ is the temperature-dependent effective number of entropy degrees of freedom. This constraint is represented by the grey band in Fig.~\ref{fig:U41_mnu4} for the reference value $m_{\rm WDM}^{\text{Ly}-\alpha}=3~\text{keV}$ but can be straightforwardly translated to a different value using Eq. (\ref{eq:bound_ly-alpha}).

\section{The Hubble tension}
\label{sec:H0}

The solution proposed in Refs.~\cite{Escudero:2019gvw,Escudero:2021rfi} to alleviate the present Hubble tension contains two key ingredients. The first is a contribution to $\Delta N_\text{eff}^\text{BBN}\sim 0.4$ that should already be present during Big Bang Nucleosynthesis. The second ingredient is an interaction rate between the Majoron and neutrinos that would exceed the Hubble rate between the BBN and CMB epochs. Thus, Majorons will thermalize with neutrinos for temperatures close to the Majoron mass $T\sim m_J$ by decay and inverse decay processes $\bar \nu_i \nu_i \leftrightarrow J$. After becoming non-relativistic, Majorons would subsequently decay into neutrinos, resulting in a slight increase of $\Delta N_\text{eff}$. In addition to this extra late radiation component, Majoron-neutrino interactions cause a damping of the neutrino free streaming by suppressing their anisotropic stress and therefore affect the determination of the Hubble constant from the CMB.

\subsection{Majoron contribution to $\Delta N_\text{eff}$}
\label{sec:MajoronDeltaNeff}

\begin{figure*}[t!]
\begin{center}
\includegraphics[width=0.7\textwidth]{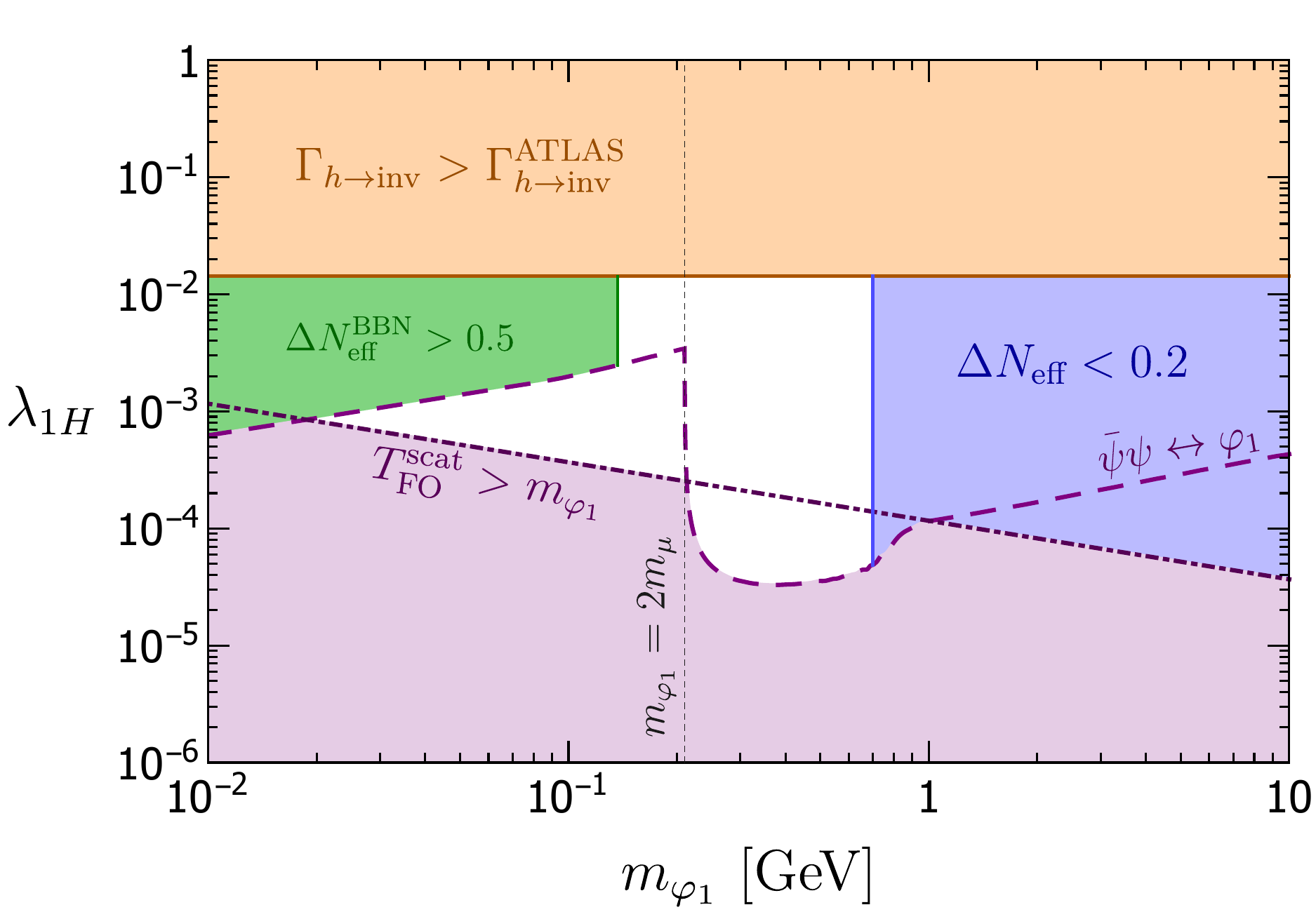}

\caption{Parameter space allowing to alleviate the Hubble tension (white). 
The blue region corresponds to values than cannot alleviate substantially the Hubble tension, while the green area represents BBN constraints on $\Delta N_\text{eff}$ estimated in Ref.~\cite{Escudero:2021rfi}. The orange region is excluded by constraints from ATLAS~\cite{ATLAS:2020kdi} on Higgs invisible decay as detailed in Sec. \ref{sec:Higgsinvisible}. The dashed purple line is determined from the (inverse) decay thermalization critera of Eq. (\ref{eq:theqdecay}) and the dash-dotted purple line corresponds to $m_{\varphi_1}<T_\text{FO}^\text{scat}$ with $T_\text{FO}^\text{scat}$ obtained from Eq. (\ref{eq:TFO}). The purple region represents the parameter space beyond the range of validity of the analysis described in Sec. \ref{sec:MajoronDeltaNeff} to determine the contribution to $\Delta N_\text{eff}$ and for which a more elaborated estimate should be performed. }
\label{fig:lambda1HVSmvarphi1}
\end{center}
 \end{figure*}

Since the scalar $\varphi_1$ mixes with the SM Higgs and also couples to the Majoron, the latter will be produced both from interactions with SM fermionic states $\psi$ mediated by virtual $\varphi_1$ as well as from $\varphi_1$ decays when it is present in the bath through the following couplings, respectively:
\begin{equation}
    \mathcal{L}_{\rm eff}\,\simeq \,  \dfrac{  \lambda_{1H} m_\psi }{2m_{\varphi_1}^2 m_h^2} \big( \partial_\mu J \partial^\mu J \bar \psi \psi \big)-\dfrac{m_\psi}{v_h} \sin(\alpha_{1H}) \varphi_1 \bar \psi \psi  
    \label{eq:Majoron_fermion_int}\,.
\end{equation}
Such couplings allow to maintain Majorons thermalized with the SM plasma until the freeze-out temperature $T_\text{FO}$ below which the Majoron population decouples from the thermal bath and behaves as background radiation, potentially leading to a contribution to $\Delta N_\text{eff}$ that can alleviate the Hubble tension. Both scatterings and (inverse) decays can allow the light scalars to thermalize and are investigated in the following.
\paragraph{Thermalization via scattering}

In order to estimate the freeze-out temperature in this case, one can compare the expansion rate of the Universe to the typical momentum-exchange rate induced by the coupling from Eq.~(\ref{eq:Majoron_fermion_int}). For the process $\bar \psi(1)+\psi(2)\rightarrow J(3)+J(4)$ such rate can be expressed as
\begin{equation}
      \left. \dfrac{\delta \rho_J}{\delta t}  \right|_{\bar \psi \psi \to JJ}   \equiv \int \prod_{i=1}^4 \frac{\diff^3 \vec p_i}{(2\pi)^3 2 E_i}  E_1 f_1(\vec p_1) f_2(\vec p_2)  |{\cal A}_{\bar{\psi}\psi\rightarrow JJ}|^2 (2\pi)^4 \delta^4(p_1+p_2- p_3 -p_4 )~,
          \label{eq:collision_term_psipsitoaa}
\end{equation}
where we follow the notations and conventions from the appendices of Ref.~\cite{Gehrlein:2019iwl}. In the relativistic limit, this quantity can be estimated as
\begin{equation}
    \left. \dfrac{\delta \rho_J}{\delta t}  \right|_{\bar \psi \psi \to JJ}    \, \simeq \, \frac{155 \pi \zeta (5) \lambda_{1H}^2  m_\psi^2 }{448 m_h^4 m_{\varphi_1}^4}  T^{11}\,.
    \label{eq:momentum_exchange_rate_psipsitoaa}
\end{equation}
By comparing this quantity to the rate of energy loss induced by the Hubble expansion we can estimate the freeze-out temperature as being
\begin{align}
    T_\text{FO}^\text{scat}\,\simeq \,0.067\;\text{GeV}\;\left( \dfrac{m_{\varphi_1}}{500\;\text{MeV}} \right)^{4/5} \left( \dfrac{0.025}{\lambda_{1H}}\right)^{2/5}\,.
    \label{eq:TFO}
\end{align}
As argued in Ref. \cite{Weinberg:2013kea}, such an estimate of the freeze-out temperature is rather reliable given the large temperature dependence of Eq. (\ref{eq:momentum_exchange_rate_psipsitoaa}), which makes the expression derived in Eq. (\ref{eq:TFO}) relatively insensitive to numerical corrections of $\mathcal{O}(1-10)$.

 \paragraph{Thermalization via (inverse) decay}
A population of $\varphi_1$ could also be produced by inverse decay of SM fermions $\psi$.  Given the fact that $\Gamma_{\varphi_1 \rightarrow  JJ} > \Gamma_{\varphi_1 \rightarrow \bar \psi \psi}$, if the  $\varphi_1$ production rate  induced by inverse decay is sizable enough to ensure $\varphi_1$ thermalization with the SM plasma, the Majorons $J$ should thermalize as well. Once the scalars $\varphi_1$ become non-relativistic and their abundance is exponentially suppressed, that we estimate to be around  $z\equiv m_{\varphi_1}/T\simeq 5 $, thermal equilibrium with the SM plasma is lost and the population of $J$ also freezes-out. As the coupling between the scalar $\varphi_1$ and $\psi$ is proportional to $m_\psi$, this process would be relevant for our parameter range mostly when the $\varphi_1$ decay channel to muons is open, i.e. for $m_{\varphi_1}>2 m_{\mu}$. The decay width of $\varphi_1$ to a pair of SM fermions is given by
\begin{equation}
    \Gamma_{\varphi_1 \rightarrow \bar \psi \psi}\,=\, c_\psi \sin^2(\alpha_{1H}) \dfrac{m_\psi^2}{8\pi v_h^2} m_{\varphi_1}\left( 1 -  \dfrac{4m_\psi^2}{m_{\varphi_1}^2} \right)^{3/2}\,,
\end{equation}
where $c_\psi$ is a colour factor. As detailed in \cite{Gehrlein:2019iwl}, the Boltzmann equation relevant for $\varphi_1$ production by (inverse) decays can be expressed in term of the yield $Y_{\varphi_1}\equiv n_{\varphi_1}/s$
\begin{equation}
    \dfrac{\diff Y_{\varphi_1}}{\diff z}\,=\,\dfrac{\Gamma_{\varphi_1 \rightarrow \bar \psi \psi}}{H(z)z} \dfrac{K_1(z)}{K_2(z)} \left[Y_\varphi^{\text{th}}(z)-Y_\varphi(z) \right]\,,
\end{equation}
where $z\equiv m_{\varphi_1}/T$. $Y_\varphi^{\text{th}}(z)$ is the thermal-equilibrium expected value of the yield and $K_{1,2}(z)$ are modified Bessel functions of the second kind. The values of the coupling $\varphi_1-\psi$ allowing to reach thermal equilibrium at $z=5$ are given by the condition
\begin{equation}
    \Gamma_{\varphi_1 \rightarrow \bar \psi \psi} \left. \dfrac{Y_\varphi^{\text{th}}(z)}{H(z)z}\dfrac{K_1(z)}{K_2(z)} \right|_{z=5}  \, \simeq \, \left( \dfrac{\lambda_{1H}}{3 \times 10^{-5}}\right)^2 \left( \dfrac{m_\psi}{m_\mu} \right)^2 \left( \dfrac{500 \, \text{MeV}}{m_{\varphi_1}} \right) \, \gtrsim \, 1 \,.
    \label{eq:theqdecay}
\end{equation}
This condition has been checked numerically and yields a rather conservative constraint on the parameter $\lambda_{1H}$. For larger couplings than the benchmark point of Eq. (\ref{eq:theqdecay}), a $J$ population would thermalize with SM fermions and freeze-out at $T_\text{FO} \simeq m_{\varphi_1}/5$.

\paragraph{Contribution to $\Delta N_\text{eff}$}
In the parameter space for which the condition of Eq. (\ref{eq:theqdecay}) is satisfied, the (inverse) decay processes are more efficient than scatterings to maintain thermal equilibrium. The resulting contribution to $\Delta N_\text{eff}$ is~\cite{Gehrlein:2019iwl}
\begin{equation}
     \Delta N_\text{eff} \,\simeq\,  0.29\,\left( \dfrac{g_{\star s}(m_\mu)}{g_{\star s}(T_\text{FO})}  \right)^{4/3} \,,
\end{equation}
with $g_{\star s}(m_\mu)\simeq 17.6$. The $\Delta N_\text{eff}$ range found in~\cite{Escudero:2019gvw,Escudero:2021rfi} that alleviates the Hubble tension is between $0.2$ and $0.5$, with a preferred value of $0.37$. In Fig.~\ref{fig:lambda1HVSmvarphi1} we depict the values of $\lambda_{1H}$ and $m_{\varphi_1}$ that would lead to such a contribution bounded by the blue line corresponding to $\Delta N_\text{eff}=0.2$ and the green line corresponding to $\Delta N_\text{eff}=0.5$. The white region represents the parameter space that allows to alleviate the Hubble tension. We emphasize that the boundaries of this white region of parameter space might be subject to small corrections given the order of magnitude estimate presented in this section. Nevertheless, a region with $0.2 < \Delta N_\text{eff} < 0.5$ will be present in that area, since we verified that at least one of the two processes analyzed would allow Majorons to thermalize down to the temperature required. In particular, in Fig. \ref{fig:lambda1HVSmvarphi1}, in the region between $2 m_\mu < m_{\varphi_1} < 700$~MeV the (inverse) decays of $\varphi_1$ allow them to thermalize with the SM bath and Majorons as long as $\lambda_{1H}$ is above the dashed purple line. Hence, the vertical boundaries from the green and blue areas correspond to when we estimate that $\varphi_1$ becomes Boltzmann-suppressed and decouples so that also the Majorons freeze-out with $0.2 < \Delta N_\text{eff} < 0.5$. Conversely, in the white triangle below the dashed purple line, scatterings with SM fermions are able to keep the Majorons in equilibrium instead with a final contribution to $\Delta N_\text{eff}$ in the same range. In some areas of the parameter space both processes may be relevant simultaneously but such a detailed analysis is beyond the scope of this work and should not lead to sizable deviations from Fig.~\ref{fig:lambda1HVSmvarphi1}.

\subsection{Majoron interactions with neutrinos}

The authors of Ref.~\cite{Escudero:2021rfi} define an effective width normalized such that for $\Gamma_\text{eff} \gtrsim 1$ Majorons do thermalize with the active neutrinos as required to alleviate the Hubble tension:

\begin{equation}
    \Gamma_\text{eff}\,\equiv\,\left( \dfrac{\lambda_\nu}{4\times 10^{-14}    } \right)^2 \left( \dfrac{0.1\;\text{eV}}{m_J} \right)\,,
    \label{eq:def_Gammaeff}
\end{equation}
where $\lambda_\nu$ is the dimensionless Majoron-neutrino coupling:
\begin{equation}
        \mathcal{L}\,\supset \, \dfrac{1}{2} \lambda_{\nu}  i J \bar \nu_i \gamma_5 \nu_i \,.
\end{equation}
In our setup this parameter corresponds to
\begin{equation}
    \lambda_{\nu}\,\equiv \,\sin  \beta \, \dfrac{m_{\nu_i}}{v_2}\,,
    \label{eq:lambdanu}
\end{equation}
where we have neglected the contribution from $\mu_{Ls}$ to $m_{\nu_i}$. In this approximation
\begin{equation}
     \Gamma_\text{eff}\,\simeq\,52\,\left( \dfrac{m_{\nu_i}}{0.05\;\text{eV}}\right)^2 \left( \dfrac{200\;\text{GeV}}{v_1}\right)^2 \left( \dfrac{0.3\;\text{eV}}{m_J} \right)\,.
     \label{eq:gammaeffexp}
\end{equation}
The best fit for $\Gamma_\text{eff}$ found in Ref.~\cite{Escudero:2021rfi} depends slightly on the number of active neutrinos interacting with the Majoron. Indeed, notice from Eq.~(\ref{eq:lambdanu}) that $\lambda_\nu$ is proportional to the neutrino mass. Thus, if the lightest neutrino is very light or massless, for instance if only two $N_R$-$N_L$ pairs are considered, its coupling to the Majoron would be negligible. In particular the best fit changes from $\Gamma_\text{eff}=67.6$ to $\Gamma_\text{eff}=59.9$ when 2 or 3 neutrinos are considered to interact with the Majoron respectively. The dependence of $\Gamma_\text{eff}$ with $\Delta N_\text{eff}^\text{BBN}$ was found to be stronger. Indeed, the best fit $\Gamma_\text{eff}=67.6$ corresponding to $\Delta N_\text{eff}^\text{BBN} = 0.37$ jumped to $\Gamma_\text{eff}=678$ for $\Delta N_\text{eff}^\text{BBN} = 0.48$, although this larger contribution to $N_\text{eff}^\text{BBN}$ significantly worsened the fit. In all cases the best fit for the Majoron mass was $m_J\sim0.3$~eV. The preferred values of $\Delta N_\text{eff}^\text{BBN} = 0.37$ and $\Gamma_\text{eff} \sim 60$ can easily be achieved as shown in Fig.~\ref{fig:lambda1HVSmvarphi1} and Eq.~(\ref{eq:gammaeffexp}).

\subsection{Constraints from Higgs invisible decay}
\label{sec:Higgsinvisible}
A coupling between the Higgs and the light scalars $J, \varphi_1$ is generated via mixing from the kinetic terms of $\phi_1$ and the $\lambda_{1H}$ term of the scalar potential as  
\begin{equation}
     \mathcal{L}\,\supset\, \sin ( \alpha_{1H}) \dfrac{h}{v_1}  \Big( \partial_\mu J \partial^\mu J \Big)-\dfrac{\lambda_{1H}}{4}v_H h \varphi_1^2\,.
     \label{eq:couplingHiggsMajorons}
\end{equation}
Via these couplings, the Higgs can decay invisibly to a Majoron or $\varphi_1$ pair with a decay rate
\begin{equation}
 \Gamma_{h\rightarrow  \text{inv}} \,=\, \Gamma_{h\rightarrow  \varphi_1 \varphi_1} + \Gamma_{h\rightarrow  J J }\,\simeq\,\dfrac{1}{64\pi} \lambda_{1H}^2  \dfrac{v_H^2}{m_h}\,,
\end{equation}
where we replaced the mixing angle $\alpha_{1H}$ by its analytical approximation in the limit of small mixing. The invisible branching ratio of the Higgs is constrained to be  $\mathcal{B}(h\to \text{inv}) < 0.11 \,(0.19)$ from ATLAS~\cite{ATLAS:2020kdi}  (CMS~\cite{Sirunyan:2018owy})
 which translates into
\begin{equation}
  \lambda_{1H} \, < \, \lambda_{1H}^\text{ATLAS} \,\simeq\,0.014 \,.
  \label{eq:constraintsInvisibleDecay}
\end{equation}

\section{Summary of the available parameter space}
\label{sec:summary}
\renewcommand*{\arraystretch}{1.2}

Taking into account all the constraints discussed in the previous sections, we sketch in Tab.~\ref{tab:parameters} the ranges for the parameters of the model in which all conditions may be satisfied so that the correct neutrino masses and mixings and dark matter relic density can be recovered together with an improvement of the Hubble tension.

Neutrino masses are controlled by the product $\theta^2 \mu_{LL}$. The parameter $\theta$ represents the mixing between the active neutrinos and the heavy pseudo-Dirac pairs and is bounded to be $\theta \leq 10^{-2}$ from tests of the PMNS unitarity via precision electroweak and flavour observables~\cite{Fernandez-Martinez:2016lgt}. Conversely, if $\theta$ is too small, the heavy pseudo-Dirac pairs, which populate the DM abundance via their decays, would not thermalize. This fixes the range for this parameter between roughly $10^{-4}$ and $10^{-2}$. We have shown in Fig.~\ref{fig:U41_mnu4} that the correct relic abundance can be obtained for $\theta = 10^{-2}, 10^{-3}$, but it can also be recovered for smaller values of $\theta$.
The parameter $\mu_{LL}$ should then correspond to $m_{\nu_i}/\theta^2$, with values in the keV to MeV range. The value of $\mu_{LL}$ in turn comes from the breaking of $L$ by two units of the vev of $\phi_2$, $v_2$. Thus, assuming order one Yukawas, $v_2$ and $\mu_{LL}$ will have a similar range as reflected in Tab.~\ref{tab:parameters}. Finally, $v_2$ is induced by the vev of $\phi_1$, $v_1$, through the $\eta$ cubic coupling so that $v_2 \simeq \eta v_1^2/m_{\varphi_2}^2$. The most natural choice for these parameters is to assume that $v_1$ and $m_{\varphi_2}$ are close to the electroweak scale, so as to avoid hierarchy problems. Thus, the suppression in $v_2$ stems from the relative smallness of $\eta$, since this parameter is protected by the additional $U(1)$ symmetry which is gained when this parameter together with $\mu_{Ls}$ and $\mu_{Rs}$ are set to zero.   

Regarding the generation and properties of DM, the most stringent constraint is on the parameter $\mu_{Ls}$. Indeed, the mixing of DM with the active neutrinos $U_{\alpha4} \sim \theta \mu_{Ls}/m_{\nu_4}$ induces its decay to X-rays, for which stringent limits exist as shown in Fig.~\ref{fig:U41_mnu4}. In particular, for $m_{\nu_4} \sim 10$~keV, $\mu_{Ls}$ is constrained to be between the eV and keV scales, depending on the value of $\theta$. On the other hand, the DM relic abundance is controlled by $\mu_{Ls}$ and $\mu_{Rs}$ and a value around 1~keV is required. Thus, as shown in Fig.~\ref{fig:U41_mnu4}, either $\mu_{Rs}$ is significantly larger than $\mu_{Ls}$, or $\theta \leq 10^{-4}$. In addition, the DM abundance is induced by the decay of the heavy neutrinos to DM and some scalar degree of freedom. Therefore, the mass of the heavy neutrinos $m_N$ should not be much higher than the TeV scale to avoid further suppressing the DM relic abundance. For large $\theta$, these neutrinos could be searched for at colliders. 

Finally, in order to alleviate the Hubble tension two main ingredients are necessary. The first is a sufficient contribution to $\Delta N_\text{eff}$ from the freeze-out of the Majorons. The main parameters controlling this are the mass of $\varphi_1$, $m_{\varphi_1}$ and its coupling to the Higgs $\lambda_{1H}$. Indeed, the Majoron is mainly aligned with the angular component of $\phi_1$ and it can be kept in thermal equilibrium most efficiently via the mixing of $\varphi_1$ with the Higgs. In order to reach $\Delta N_\text{eff} \sim 0.4$, the Majoron must decouple roughly with the muons, which can happen for $m_{\varphi_1} < 1$~GeV, as shown in Fig.~\ref{fig:lambda1HVSmvarphi1}. Regarding the coupling, the lack of evidence for an invisible Higgs decay at LHC requires $\lambda_{1H} \lesssim 0.01$. On the other hand, $\lambda_{1H} \geq 10^{-4}$ is necessary to keep the Majorons in thermal equilibrium until sufficiently late times. The second ingredient required to alleviate the Hubble tension is a coupling between the Majoron and the active neutrinos that allows them to thermalize after BBN and a Majoron mass around the eV scale so that it will decay to neutrinos and contribute to $\Delta N_\text{eff}$ at CMB. This decay width depends on the ratio of the neutrino masses over $v_1$, as well as on the mass of the Majoron itself and the correct value is obtained for $v_1 \sim 100$~GeV for $m_J \sim 1$~eV.

\begin{table}[t]
\centering
\begin{tabular}{|c||c|c|c|c|c|}
\hline \textbf{Parameter} & $m_{\nu_4}$ &  $m_{N}$ & $m_{\varphi_1}$  & $m_{\varphi_2}$ & $v_1$  \\
\hline \textbf{Range} & $[10,10^3]$ keV & $[10^2,10^3]$ GeV  &$[10^{-1},1]$ GeV & $[10,10^3]$ GeV & $[1,10^{3}]$ GeV   \\ 
\hline
\hline \textbf{Parameter}  &  $v_2$ & $\eta$ & $\mu_{ab} (a,b=R,L,s)$  & $\theta$ &  $\lambda_{1H}$   \\
\hline \textbf{Range}  & keV-MeV & keV-MeV &  keV-MeV & $[10^{-4},10^{-2}]$ &  $[10^{-4},10^{-2}]$   \\
\hline
\end{tabular}
\caption{Order of magnitude for the allowed range of some relevant parameters allowing to simultaneously explain neutrino masses, the dark matter relic abundance and alleviate the Hubble tension.}
\label{tab:parameters}
\end{table}

\section{Conclusions}
\label{sec:conclusions}

Extending the Standard Model particle content with right-handed neutrinos is arguably the simplest extension able to account for the evidence of neutrino masses and mixings. In order to also provide a natural explanation to the smallness of neutrino masses, two options emerge. In the canonical, high-scale type-I Seesaw the ratio between the electroweak scale and the large Majorana mass provides naturally the required suppression. Conversely, in low-scale realizations, such as the linear or inverse Seesaw, the Lepton number $L$ symmetry that protects neutrino masses is instead exploited. If this symmetry is approximate and only broken by small parameters, these will also naturally suppress the generation of neutrino masses. These low-scale realizations have the twofold advantage of enhancing the relevant phenomenological impact of the model and hence its testability, as well as avoiding a contribution to the Higgs hierarchy problem. 

We have explored the possibility that the small breaking of the $L$ symmetry in the inverse Seesaw is dynamical. Its smallness emerges from a Seesaw-like structure in the scalar sector in which the vev of the $L=2$ scalar responsible for neutrino masses is only indirectly induced by a vev around the electroweak scale, as in the type-II Seesaw. The parameter linking the two is small in a technically natural way since it is protected by an additional symmetry. 

This spontaneous breaking of $L$ in turn leads to the existence of a Majoron. We have explored the parameter space and conclude that this Majoron may contribute to the number of relativistic degrees of freedom in the early Universe as well as couple to the active neutrinos with the required values as to significantly alleviate the Hubble tension. This possibility is mainly constrained by the invisible decay of the Higgs, since the Majoron production critically depends on the mixing between the new scalar that breaks the $L$ symmetry and the Higgs. Nevertheless, an order of magnitude smaller mixings than presently allowed by LHC constraints would still allow for a solution to the Hubble tension.

Among the new neutrinos introduced in low-scale Seesaws, two options exist due to the approximate $L$ symmetry. The first are pseudo-Dirac pairs in which the left-handed component has a sizable mixing with the active neutrinos. The second are Majorana sterile neutrinos with couplings suppressed by the $L$-breaking parameters. For keV-scale masses, these Majorana sterile neutrinos may be sufficiently stable to be good dark matter candidates. While production via mixing through the Dodelson-Widrow mechanism is excluded by X-ray searches, we have shown that the correct relic abundance may be obtained for appropriate values of the model parameters via the freeze-in decays of the heavier pseudo-Dirac pairs to the new scalars and dark matter. These same couplings also control the mixing of the active neutrinos with dark matter as well as with the heavy pseudo-Dirac pairs. The main constraints on these mixings come from searches of the dark matter decays to X-rays and from unitarity tests of the PMNS mixing matrix from precision electroweak and flavour observables respectively. While the combination of these two probes rules out significant parts of the allowed parameter space, the correct relic abundance can still be obtained from the parameter that controls the mixing of dark matter with the right-handed component of the pseudo-Dirac pair. This mixing is more difficult to constrain, since the SM active neutrinos mainly mix to the left-handed component. Two possibilities are then viable. If the mixing between the active neutrinos and the heavy pseudo-Dirac pairs is sizable, close to their PMNS unitarity constraints, then the parameter that controls the dark matter-right-handed neutrino mixing needs to be significant. This implies a hierarchy of four or five orders of magnitude with respect to the mixing with the left-handed component, which may be considered fine-tuned. Conversely, if the two couplings are similar, the mixing of the heavy neutrinos with the active states needs to be very suppressed, reducing the testability of the model through PMNS unitarity deviations and, eventually, direct production at colliders.

Finally, the heavy pseudo-Dirac pairs could possibly explain the baryon asymmetry of the Universe through the ARS baryogenesis via leptogenesis mechanism~\cite{Akhmedov:1998qx,Asaka:2005pn,Shaposhnikov:2008pf,Canetti:2010aw,Abada:2015rta,Hernandez:2015wna,Hernandez:2016kel,Drewes:2017zyw,Abada:2017ieq,Chun:2017spz,Abada:2018oly,Caputo:2018zky}. While we have assumed that the heavy pseudo-Dirac neutrinos thermalize, the ARS leptogenesis mechanism requires that at least some of them do not reach thermal equilibrium. This could be an option, since we only require one of them to thermalize in order to populate the DM abundance via its decays. Moreover, the correct DM density might also be obtained without thermalization of the heavy pseudo-Dirac pairs. This possibility together with the impact of the additional interactions of the heavy pseudo-Dirac pairs in the context of leptogenesis would be an interesting extension of the present study.

To summarize, we have shown that the SM extension considered with a dynamical breaking of the $L$ symmetry characterizing the inverse Seesaw, is able to account simultaneously for the observed neutrino masses and mixings in a natural way as well as to provide a dark matter candidate with the correct relic abundance and alleviate the present Hubble tension between CMB and supernovae observations. The main constraints on the allowed parameter space come from unitarity tests of the PMNS mixing matrix through precision electroweak and flavour observables, searches for invisible Higgs decays at the LHC and X-ray searches for this decay mode of the sterile neutrino dark matter candidate.

\section*{Acknowledgments}
We warmly thank Pilar Coloma for very helpful discussions and for participating in the initial phases of the project. We are also very grateful to Asmaa Abada, Giorgio Arcadi, Miguel Escudero, Pilar Hernandez, Jacobo Lopez-Pavon and Michele Lucente for very illuminating discussions. The authors acknowledge support from the Spanish Agencia Estatal de Investigaci\'{o}n and the EU ``Fondo Europeo de Desarrollo Regional'' (FEDER) through the projects  FPA2015-65929-P,  PID2019-108892RB-I00/AEI/10.13039/501100011033, PGC2018-095161-B-I00, and Red Consolider MultiDark FPA2017-90566-REDC. The authors are also supported by the IFT Centro de Excelencia Severo Ochoa Grant SEV-2016-0597.
This project has received support from the European Union’s Horizon 2020 research and innovation programme under the Marie Sk\l odowska -Curie grant agreement No 860881-HIDDeN.

\bibliographystyle{JHEP}
\bibliography{Majoron}

\end{document}